\documentstyle[letter,mbf,wrapft]{ptptex}
\newcommand{\dN}{\Delta N}
\newcommand{\G}{\Gamma}

\newcommand{\Xd}{X^\dagger}

\newcommand{\Xbn}{\bar{X}}
\newcommand{\Xbd}{\bar{X}^\dagger}
\newcommand{\Bb}{\bar{B}}

\newcommand{\ket}[1]{\left| {#1}\,\right\rangle}
\newcommand{\bra}[1]{\left\langle {#1}\right|}
\newcommand{\braket}[2]{\left\langle {#1}|{#2}\right\rangle}
\newcommand{\kket}[1]{\left.\left| {#1}\,\right\rangle\!\right\rangle}
\newcommand{\bbra}[1]{\left\langle\!\left\langle {#1}\right|\right.}
\newcommand{\bbrakket}[2]{\left\langle\!\left\langle {#1}|{#2}\right\rangle\!\right\rangle}
\newcommand{\ua}{u_\alpha}
\newcommand{\va}{v_\alpha}
\renewcommand{\aa}{a_\alpha}
\newcommand{\ata}{a_{\tilde{\alpha}}}
\newcommand{\bracket}[3]{{\left\langle {#1} \, \right|}{ {#2} \left| {#3} \,\right\rangle}}
\notypesetlogo  
\title{
Intrinsic Pair Operators and the Number-conserved Treatment of Pairing Correlations in Nuclei
}

\author{
Makoto {\sc Ueno}\footnote{Present address, FUJISOFT ABC Inc.,
        6-26, Nishiki 1-chome, Naka-ku, Nagoya 460-0003.}
, Daisuke {\sc Hayashi}
and Yoshinao {\sc Miyanishi}
}

\inst{
Graduate School of Science, Nagoya University, Nagoya 464-8602
}

\recdate{
}

\abst{
By introducing the intrinsic pair operators which commute with number 
fluctuation operator, a new formalism is given for the number-conserving 
description of the pairing correlations. The difficulty in the conventional 
RPA treatment for pairing rotation is removed. 
}
\begin{document}
\maketitle
It is well known that the pairing correlations play an important role 
in various phenomena in nuclei\cite{rf:R-S}. 
The correlations are usually described by the BCS theory or 
Hartree-Fock-Bogoliubov (HFB) theory. 
In a super phase the particle number is conserved only in average and
its fluctuation is important in a finite system such as nuclei \cite{rf:N-FLU}. 

The projection method is the rigorous one to restore the particle number 
conservation \cite{rf:N-PRJ}. 
However it loses the simple picture of excitation modes. 
On the other hand the RPA method treats the number fluctuation mode as 
a zero-energy eigen mode and recovers the broken symmetry within 
its approximation \cite{rf:RPA} .
However it has some difficulties related with the small amplitude approximation. 
For example, the norm of the RPA ground state becomes divergent. 
Extension of RPA has been made in various ways although such an approach
becomes rather complicated \cite{rf:M-RPA}. 

In this letter we propose a new  method to treat the number conservation. 
We first define the intrinsic pair modes which commute with the particle number 
operator and extract the intrinsic Hamiltonian made up of only these modes.
The wave function is formally given by the number projection on the intrinsic
sates. 
However the integral calculation for the projection is not necessary, 
because the eigen value problem of the original Hamiltonian reduce to that
of the intrinsic Hamiltonian.
Thus we take account of both merits of RPA and the projection method. 

We start with a general effective Hamiltonian,
\begin{equation}
H=\sum_\alpha \epsilon_\alpha c_\alpha^\dagger c_\alpha
 +\sum_{\alpha\beta\gamma\delta} v_{ \alpha\beta\gamma\delta}
  c_\alpha^\dagger c_\beta^\dagger c_\delta c_\gamma, 
\end{equation}
where $c_\alpha^\dagger$ and $c_\alpha$ are the creation and annihilation 
operator for the single particle state with the quantum number $\alpha$
and $\epsilon_\alpha$ denotes the single particle energy.
For simplicity we discuss the system with one kind of nucleon 
(proton or neutron) and the BCS case in the following.

After the Bogoliubov-Valatin transformation,
\begin{equation}
\aa^\dagger = \ua c_\alpha^\dagger-\va c_{\tilde{\alpha}},\\
\end{equation}
the particle number operator $N=\sum_\alpha c_\alpha^\dagger c_\alpha$
is written by the quasi-particle operators as,
\begin{equation}
N=\langle N \rangle+N_c+N_s,
\end{equation}
\begin{equation}
N_c = \sum_\alpha \ua \va (\aa^\dagger\ata^\dagger+\ata\aa),\quad
N_s = \sum_\alpha (\ua^2-\va^2)\aa^\dagger\aa,
\end{equation}
where $\tilde{\alpha}$ denotes the time reversal of $\alpha$. 
$\langle N \rangle$ represents the expectation value of $N$
for the BCS vacuum $ \ket{0}$. 
The number fluctuation operator is defined as
\begin{equation}
\dN\equiv N-\langle N \rangle = N_c+N_s. 
\end{equation}
$N_c$ is rewritten as
\begin{equation}
N_c=L\,(\Xd_c+X_c),\quad L^2 \equiv \langle \dN^2 \rangle = 
                         2 \sum_\alpha u_\alpha^2 v_\alpha^2 
\end{equation}
by introducing the Tamm-Dankoff (TD) type collective operator $\Xd_c$, 
\begin{equation}
\Xd_c=\sum_{\alpha\beta}\psi_c(\alpha\beta) 
            \frac{\aa^\dagger a_\beta^\dagger}{\sqrt{2}},
\end{equation}
\begin{equation}
\psi_c(\alpha\beta) =\frac{\sqrt{2}\ua\va}{L}\delta_{\beta\tilde{\alpha}},
\quad \sum_{\alpha\beta}\psi_c^2(\alpha\beta)=1.
\end{equation}

In super phase $L$ is assumed to be large so that we regard $\varepsilon\equiv 1/L$ as 
a small parameter. 
Noticing that $\bra{0} X_c\Xd_c \ket{0} =1 $, $X_c$ and $\Xd_c$ 
are regarded as a quantity of order $\varepsilon^0$ and therefore 
$N_c$ is $O(\varepsilon^{-1})$. 
On the other hand
\begin{equation}
\bra{0} X_c N_s \Xd_c \ket{0}
=\sum_\alpha (\ua^2-\va^2)\psi_c^2(\alpha\tilde{\alpha})
=1-2\sum_\alpha\va^2\psi_c^2(\alpha\tilde{\alpha})
<1
\end{equation}
so that $N_s$ is at most $O(\varepsilon^0)$. 

Let us define the  TD type non-collective mode as
\begin{equation}
\Xd_n=\sum_{\alpha\beta}\psi_n(\alpha\beta)
      \frac{\aa^\dagger a_\beta^\dagger}{\sqrt{2}},
\end{equation}
which satisfies
\begin{equation}
\langle 0| X_c\Xd_n |0 \rangle=0,
\qquad\langle 0| X_n\Xd_{n'} |0 \rangle=\delta_{nn'}. 
\end{equation}
This leads to the following equation for $\psi_n(\alpha\beta)$
\begin{equation}
\sum_{\alpha\beta}\psi_c(\alpha\beta)\psi_n(\alpha\beta)=0,
\qquad
\sum_{\alpha\beta}\psi_n(\alpha\beta)\psi_{n'}(\alpha\beta)=\delta_{nn'}. 
\end{equation}
Such a $\psi_n(\alpha\beta)$ can be obtained, for example, from the
eigen equation for stability matrix,
\begin{equation}
\sum_{\gamma\delta} ({\cal A}_{\alpha\beta ;\gamma\delta}
                  + {\cal B}_{\alpha\beta ;\gamma\delta}) \psi_n(\gamma\delta)
= \omega_n \psi_n(\alpha\beta) \qquad (\omega_n >0),
\end{equation}
where
\begin{equation}
{\mathcal A}_{\alpha \beta ;\gamma \delta}= \bracket{0}{[a_{\beta}a_{\alpha},[\hat H,a^{\dagger}_{\gamma} a^{\dagger}_{\delta}]]}{0},\qquad {\mathcal B}_{\alpha \beta ;\gamma \delta}= \bracket{0}{[[\hat H,a_{\alpha}^{\dagger} a_{\beta}^{\dagger}], a_{\delta}^{\dagger}a_{\gamma}^{\dagger}]}{0}
\end{equation}
It is well known the eigenvector with zero eigenvalue corresponds to 
$\psi_c(\alpha\beta)$. 
It should be mentioned that the wave functions $\psi_n(\alpha\beta)$ thus determined are not unique and there remains the freedom of unitary transformation between them. 
However they span the complete set together with $\Xd_c$ in the sense that
\begin{equation}
\frac{a_\alpha^\dagger a_\beta^\dagger}{\sqrt{2}}
= \sum_\mu\psi_\mu(\alpha\beta)\Xd_\mu,
\end{equation}
where we use the notation $\mu=c$ or $n$.

Now we define the intrinsic pair operators $\Xbd_n$ and $\Bb_q$ corresponding 
to $\Xd_n$ and $B_q \equiv a_\alpha^\dagger a_\beta$ by the following equations;
\begin{equation}
[\dN,\Xbd_n]=0
\qquad
\Xbn_n\ket{0}=0, 
\end{equation}
\begin{equation}
[\dN,\Bb_q]=0,
\qquad
\Bb_q\ket{0}=0. 
\end{equation}
Here we assume that they are obtained by an expansion form,
\begin{equation}
\label{eq:Xbd}
\Xbd_n=\Xd_n
+\varepsilon\Xbd_n\mbox{}^{(1)}+\varepsilon^2\Xbd_n\mbox{}^{(2)}+ \cdots,
\end{equation}
\begin{equation}
\label{eq:Bb}
\Bb_q=\Bb_q^{(0)}
+\varepsilon\Bb_q^{(1)}+\varepsilon^2\Bb_q^{(2)}+ \cdots. 
\end{equation}
The leading term $\Bb_q^{(0)}$ is given by
\begin{equation}
\Bb_q^{(0)}=B_q
-\G_{cc}(q)\Xd_c X_c
-\sum_n\left[\G_{cn}(q)\Xd_n X_c+\G_{nc}(q)\Xd_c X_n\right], 
\end{equation}
where $\G_{\mu\nu}(q)$ is defined as 
\begin{equation}
\G_{\mu\nu}(q)=2\sum_\gamma \psi_\mu(\alpha\gamma) \psi_\nu(\beta\gamma).
\end{equation}
Note that $[X_c,\Bb_q]=0$ and $[\Xd_c,\Bb_q]=0$ in lowest order.
From Eqs.(\ref{eq:Xbd}) and (\ref{eq:Bb}) the pair operators $X_n$ and $B_q$ 
are conversely expanded as
\begin{equation}
\label{eq:xexp}
\Xd_n
=\Xbd_n
+\varepsilon\Xd_n\mbox{}^{(1)}
+\varepsilon^2\Xd_n\mbox{}^{(2)}+\cdots,\\
\end{equation}
\begin{equation}
\label{eq:bexp}
B_q
=B_q^{(0)}
+\varepsilon B_q^{(1)}
+\varepsilon^2 B_q^{(2)}+\cdots,
\end{equation}
where $\Xd_n\mbox{}^{(i)}$ $(i\geq1)$, $B_q^{(i)}$ $(i\geq 0)$ are expressed by 
a function of $\Xbd_\mu$, $\Xbn_\mu$ and $\Bb_q$. 
Since the Hamiltonian is expressed by the pair operators $\Xd_\mu$, $X_\mu$ 
and $B_q$, we can rewritten it by the intrinsic pair operators and $\dN$. 
\begin{equation}
\label{eq:hsep}
H=\langle H \rangle+H_{rot}+H_{intr}+H_{coupl}. 
\end{equation}
Here $H_{rot}$ is a function of $\dN$
\begin{subequations}
\begin{eqnarray}
H_{rot}&=&E_{rot}+C_1\dN+C_2\dN^2+\cdots,\\
&&E_{rot}=-C_2\langle\dN^2\rangle-\cdots. 
\end{eqnarray}
\end{subequations}
$H_{intr}$ is composed of only the intrinsic pair operator, 
\begin{equation}
H_{intr}=H_{intr}^{(0)}+\varepsilon H_{intr}^{(1)}+\varepsilon^2 H_{intr}^{(2)}+\cdots,
\end{equation}
the leading term of which is given as
\begin{equation}
\label{eq:hintr0}
H_{intr}^{(0)}
=\sum_\alpha E_\alpha \Bb_{\alpha\alpha}
+\sum_{nn'}V_X(nn')\Xbd_n\Xbn_{n'}
+\sum_{nn'}V_V(nn')\left(\Xbd_n\Xbd_{n'}+h.c.\right),
\end{equation}
\begin{subequations}
\begin{eqnarray}
V_X(nn')&=&\sum_{\alpha\beta\gamma\delta}
{\cal A}_{\alpha\beta;\gamma\delta}\psi_n(\alpha\beta)\psi_{n'}(\gamma\delta)
\nonumber\\
&&-\sum_{\alpha\beta}(E_\alpha+E_\beta)\psi_n(\alpha\beta)\psi_{n'}(\alpha\beta),\\
V_V(nn')&=&\sum_{\alpha\beta\gamma\delta}
{\cal B}_{\alpha\beta;\gamma\delta}\psi_n(\alpha\beta)\psi_{n'}(\gamma\delta),
\end{eqnarray}
\end{subequations}
where $E_\alpha$ denotes the quasi-particle energy.
$H_{coupl}$ is a function of $\dN$ and intrinsic pair operators. 
It should be noticed that $\Xd_c$ and $X_c$ terms in Eqs.(\ref{eq:xexp}) and
(\ref{eq:bexp}) are absorbed in $\dN$ term in Eq.(\ref{eq:hsep}), 
because the relation $[H,\dN]=0$ should be hold. 
Thus we separate the original Hamiltonian into three parts, the rotational part, intrinsic part and their coupling part. 

Now we span the intrinsic subspace by a set of intrinsic states
\begin{equation}
\ket{0},\quad
\Xbd_n\ket{0},\quad
\Xbd_n\Xbd_{n'}\ket{0},\cdots,
\end{equation}
which are named generically as
\begin{equation}
\ket{\bar{k}}=\Xbd_{n_1}\Xbd_{n_2}\cdots\Xbd_{n_l}\ket{0}. 
\end{equation}
Then the number projected intrinsic states are given by
\begin{eqnarray}
\label{eq:pk}
\kket{\bar{k}}&=&P_N\ket{\bar{k}},\qquad
P_N\equiv\int_0^{2\pi}d\phi \,\mbox{e}^{i\phi\dN}. 
\end{eqnarray}
The energy matrix and the norm matrix between these states are given by
\begin{eqnarray}
{\cal H}_{kk'}
&\equiv&\bbra{\bar{k}}H\kket{\bar{k'}}\nonumber\\
&=&\bra{0}\Xbn_{n_l}\cdots\Xbn_{n_1}P_NH\,P_N\Xbd_{n'_1}\cdots\Xbd_{n'_l}\ket{0}\nonumber\\
&=&\bra{0}P_N\ket{0}\bra{\bar{k}}H_{intr}\ket{\bar{k'}},\\
{\cal N}_{kk'}
&\equiv&\bbrakket{\bar{k}}{\bar{k'}}\nonumber\\
&=&\bra{0}\Xbn_{n_l}\cdots\Xbn_{n_1}P_N\Xbd_{n'_1}\cdots\Xbd_{n'_l}\ket{0}\nonumber\\
&=&\bra{0}P_N\ket{0}\braket{\bar{k}}{\bar{k'}}. 
\end{eqnarray}
Here we use the relation $P_N\dN=\dN\,P_N=0$ and $[P_N,\Xbd_n]=[P_N,\Xbn_n]=0$. 
Therefore the eigenvalue problem reduces to that of $H_{intr}$, i.e.,
\begin{equation}
\label{eq:diag}
\sum_{k'}\bar{\cal H}_{kk'}u_{k'}^l=E_l\sum_{k'}\bar{\cal N}_{kk'}u_{k'}^l,
\end{equation}
where $\bar{\cal H}_{kk'}=\bra{\bar{k}}H_{intr}\ket{\bar{k'}}$, $\bar{\cal N}_{kk'}=\braket{\bar{k}}{\bar{k'}}$. 
Note that $\bra{0}P_N\ket{0}$ is factored out so that we need no integration. 

Since the lowest order of $H_{intr}$ is given by Eq.(\ref{eq:hintr0}) we can diagonalized it by introducing the RPA mode in the intrinsic subspace (which we call
IRPA mode)
\begin{equation}
\bar{\xi}_\lambda^\dagger
=\sum_n \left[{\cal X}_\lambda(n)\Xbd_n-{\cal Y}_\lambda(n)\Xbn_n\right].
\end{equation}
The wave functions, ${\cal X}_\lambda(n)$, ${\cal Y}_\lambda(n)$, 
are determined so as to satisfy
\begin{equation}
[H,\bar{\xi}_\lambda^\dagger]=\omega_\lambda\bar{\xi}_\lambda^\dagger+O(\varepsilon),
\end{equation}
which gives the IRPA equation for ${\cal X}_\lambda(n)$ and ${\cal Y}_\lambda(n)$;
\begin{subequations}
\begin{eqnarray}
\omega_\lambda {\cal X}_\lambda(n)
&=&E_n {\cal X}_\lambda(n)
+\sum_{n'}V_X(nn'){\cal X}_\lambda(n')
+\sum_{n'}V_V(nn'){\cal Y}_\lambda(n'),\\
-\omega_\lambda {\cal Y}_\lambda(n)
&=&E_n {\cal Y}_\lambda(n)
+\sum_{n'}V_X(nn'){\cal Y}_\lambda(n')
+\sum_{n'}V_V(nn'){\cal X}_\lambda(n'). 
\end{eqnarray}
\end{subequations}
Thus $H_{intr}^{(0)}$ is written as
\begin{equation}
H_{intr}^{(0)}=E^{(0)}_{vib}+\sum_\lambda\omega_\lambda\bar{\xi}_\lambda^\dagger\bar{\xi}_\lambda,
\end{equation}
\begin{equation}
E^{(0)}_{vib}=-\sum_\lambda \omega_\lambda {\cal Y}_\lambda^2(n).
\end{equation}
It should be noticed that the IRPA vacuum is defined as the vacuum for only the intrinsic mode and no difficulty arises in the conventional RPA vacuum which is defined as the vacuum for all the eigen mode including the zero-energy mode. 

The higher order terms of $H_{intr}$ are given by more complicated functions of 
$\Xbd_n$, $\Xbn_n$, $\Bb_q$. 
We must generally solve the eigenvalue problem of Eq.(\ref{eq:diag}) for 
the higher order calculations. 

\begin{wraptable}{l}{7.5cm}
\caption{Ground state energies (in MeV) for Sn isotopes are shown.}
\label{tbl:sn}
\begin{tabular}{crrrrr} \hline \hline
A & BCS & RPA & PW0 & PW2 & exact\\ \hline
104 & -2.313 & -2.742 & -2.742 & -2.744 & -2.745 \\
108 & -1.844 & -2.526 & -2.526 & -2.534 & -2.534 \\
112 & 0.930 & 0.140 & 0.140 & 0.131 & 0.134 \\
116 & 6.030 & 5.216 & 5.216 & 5.193 & 5.180 \\
120 & 13.489 & 12.761 & 12.761 & 12.758 & 12.761 \\
124 & 22.903 & 22.330 & 22.330 & 22.334 & 22.333 \\
128 & 33.837 & 33.499 & 33.499 & 33.501 & 33.500 \\
 \hline
\end{tabular}
\end{wraptable}

In order to illustrate the applicability of our method we calculate the ground state
energy of the system with the constant pairing interaction. It is well known 
that the exact solution for the state with seniority 0 is easily obtained 
numerically.
The numerical results for Sn isotopes are shown in Table \ref{tbl:sn}. 
The single particle energy $\epsilon_\alpha$ and the pairing force strength 
are taken from Ref.\cite{rf:K-S}. 
We performed the calculation up to $O(\varepsilon^2)$ and refer it as PW2, while
the calculation up to $O(\varepsilon^0)$ is denoted by PW0. 
We find that the results of PW0 are completely agree with those of the conventional RPA.
This is not trivial because each value of $E_{rot}$ and $E_{vib}$ is different 
between RPA and PW0. It is also found that PW2 calculation generally improve
the RPA results.

In summary we propose a new method to treat the number conservation. 
By introducing the intrinsic pair modes which commute with the particle number 
operator the original Hamiltonian is divided into three parts; the rotational,
the intrinsic and the coupling part. 
Thereby we never treat the rotational mode as the zero energy eigen mode, 
and therefore we circumvent the difficulties lying in the conventional RPA. 
The wave function is formally projected to the states with definite particle 
number.
However we need no integration over the phase angle as in the usual projection
method because the eigen value problem is reduced to that of the intrinsic
subspace. 
The numerical calculation was done to illustrate the applicability 
of our method and the result shows that our method works well.
The detail will appear in a subsequent paper.


\end{document}